\documentclass{ws-procs9x6}

\begin{document}

\title{Variable-Mass Dark Matter and the Age of the Universe
\footnote{\uppercase{P}resented by \uppercase{R}. \uppercase{R}osenfeld at the 
\uppercase{V SILAFAE}, \uppercase{J}uly 12-16, \uppercase{L}ima, 
\uppercase{P}eru.}}

\author{U. Fran\c{c}a}

\address{SISSA / ISAS\\
Via Beirut 4 \\ 
34014 Trieste, Italy\\ 
E-mail: urbano@sissa.it}

\author{R. Rosenfeld}

\address{Instituto de F\'{\i}sica Te\'orica-UNESP\\ 
Rua Pamplona, 145 \\
01405-900, S\~ao Paulo, SP, Brazil\\
E-mail: rosenfel@ift.unesp.br}  

\maketitle

\abstracts{
Models with variable-mass particles, called VAMPs, have been proposed as a solution 
to the cosmic coincidence problem that plagues dark energy models. In this contribution 
we make a short description of this class of models and explore some of its observational
consequences. In particular, we show that fine tuning is still required in this
scenario and that the age of the Universe is 
considerably larger than in the standard $\Lambda$-CDM model.}

\section{Introduction}

The Wilkinson Microwave Anisotropy Probe (WMAP) satellite 
\cite{wmap}, along with other experiments, 
has confirmed that the universe is very nearly flat, 
and that there is some form of
dark energy (DE) that is the current dominant energy component, 
accounting for approximately 70$\%$ of the critical density. 
Non-baryonic cold dark matter (DM) contributes around 25$\%$ to the
total energy density of the universe and the remaining 
5$\%$ is the stuff we are made of, baryonic matter. 

DE is
smoothly distributed throughout the universe and its equation of state with
negative pressure is causing its present acceleration.
It is generally modelled using a scalar
field, the so-called quintessence models, either slowly rolling towards the 
minimum of the potential or already trapped in this minimum
\cite{darkenergy}. 

An intriguing possibility is that DM particles could interact with the DE
field, resulting in a time-dependent mass. In this scenario, 
dubbed VAMPs (VAriable-Mass Particles)\cite{carrolvamps}, the mass of 
the DM particles evolves according to some function 
of the dark energy field $\phi$ 
\cite{carrolvamps,quirosvamps,peeblesvamps,hoffmanvamps,exp1amendola,exp2amendola,exp3amendola,pietronivamps,riottovamps}. 
In this case, the DM
component can have an effective equation of state with negative pressure
that could present the same behaviour as DE.
 
We studied a model with exponential coupling between DM and DE, since it 
presents a tracker solution where the effective equation of state of DM 
mimics the 
effective equation of state of DE and
the ratio between DE and DM energy density remains constant after this attractor
is reached.
This behavior could solve the ``cosmic coincidence problem'', that is,
why are the DE and DM energy densities similar today.  

This type of solution also appears when the DE  
field with a exponential potential is not coupled to the other 
fluids. 
In fact, Liddle and Scherrer  \cite{tracker}
showed that for a non-coupled DE, the exponential potential is the only
one that presents stable tracker solutions. 
In this case, however, it is 
not able to explain the current acceleration of the universe and other observational constraints,
unless we assume that the field has not yet reached the 
fixed point regime \cite{jhep}. 

In this contribution, we review our results presented in \cite{prd_us}.

\section{A simple exponential VAMP model} \label{sec:idea}

In the exponential VAMP model, the potential of 
the DE scalar field $\phi$ is given by
\begin{equation} \label{eq:potential}
V(\phi)  = V_0 \ e^{\beta \phi/ m_p} , 
\end{equation}
where $V_0$ and $\beta$ are positive constants and 
$m_p = M_p / \sqrt{8 \pi} = 2.436 \times 10^{18}$ GeV is the reduced Planck mass
in natural units, $\hbar = c = 1$. 
Dark matter is modelled by a scalar particle $\chi$ of mass 
\begin{equation} \label{eq:vampmass}
M_{\chi} = M_{\chi 0} \ e ^{- \lambda (\phi- \phi_0) /m_p} \ ,
\end{equation}
where $M_{\chi 0}$ is the current mass of the dark matter particle  (hereafter the index $0$ denotes
the present epoch, except for the potential constant $V_0$) and $\lambda$ is a
positive constant. 

In this case we can show that
\begin{eqnarray} \label{eq:bothfluids}
\dot{\rho}_{\chi} + 3 H \rho_{\chi} (1 + \omega^{(e)}_{\chi}) = 0 \ , \\
\dot{\rho}_{\phi} + 3 H \rho_{\phi} (1 + \omega^{(e)}_{\phi}) = 0 \ ,
\end{eqnarray}
where
\begin{equation} \label{eq:vampeffstate}
\omega^{(e)}_{\chi} =\frac{\lambda \dot{\phi}}{3 H m_p} = \frac{\lambda \phi'}{3 m_p}\ ,
\end{equation}
\begin{equation} \label{eq:deeffstate}
\omega^{(e)}_{\phi} = \omega_{\phi} -\frac{\lambda \dot{\phi}}{3 H m_p} \ \frac{\rho_{\chi}}{\rho_{\phi}}
=\omega_{\phi} -\frac{\lambda \phi'}{3 m_p} \ \frac{\rho_{\chi}}{\rho_{\phi}}  \ ,
\end{equation}
are the effective
equation of state parameters for dark matter and dark energy, respectively. 
Primes denote derivatives with respect to $u=  \ln (a) = - \ln (1+z)$,
where $z$ is the redshift, and $a_0 = 1$.

At the present epoch the energy density of the universe is divided 
essentially between
dark energy and dark matter. In this limit, there is a fixed point solution for
the equations of motion of the scalar field such that 
\begin{equation} \label{eq:fpOmega}
\Omega_{\phi} = 1 - \Omega_{\chi} = \frac{3}{(\lambda + \beta)^ 2} + \frac{\lambda}{\lambda + \beta} 
\end{equation}
\begin{equation} \label{eq:fpomega}
\omega^{(e)}_{\chi} = \omega^{(e)}_{\phi} = - \ \frac{\lambda}{\lambda + \beta}.
\end{equation}

The equality between $\omega_{\chi}$ and $\omega_{\phi}$ in the attractor regime
comes from the tracker behavior of the exponential potential 
\cite{exp1amendola,riottovamps} in this regime. 

\begin{figure}[ht]
\vspace{0.5cm}
\includegraphics[scale=0.4,angle=90]{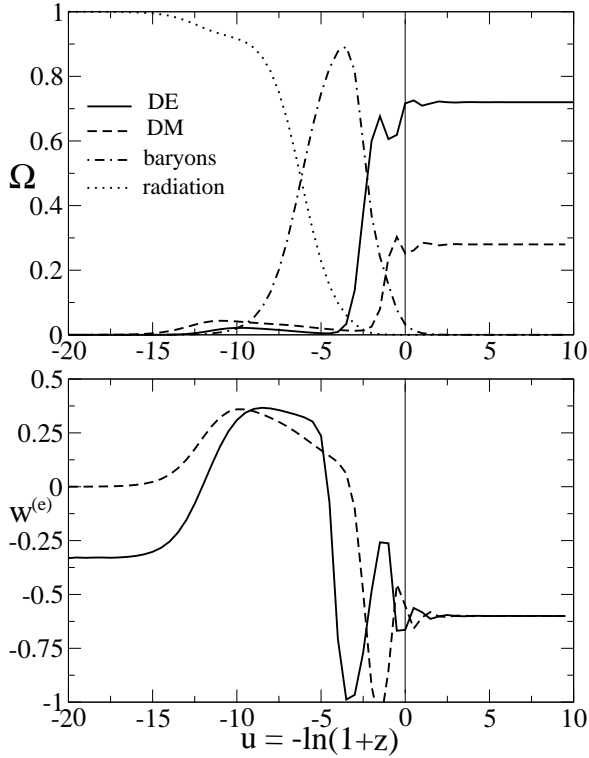}
\caption{\label{fig:omegas} \small {\it Top panel}: Density parameters of the 
components of the universe
as a function of $u=- \ln(1+z)$ for $\lambda = 3$, $\beta = 2$ and 
$V_0 = 0.1 \widetilde{\rho_c}$. 
After a transient period of baryonic matter domination (dot-dashed line), DE comes
to dominate and the ratio between the DE (solid line) and DM (dashed line) energy densities 
remains constant.
{\it Bottom panel}: Effective equations of state for DE (solid line) and DM (dashed line)
for the same parameters used in top panel. In the tracker regime both equations of state 
are negative.}
\end{figure}

However, this is only valid in the fixed point regime. If we want to know what
happens before, the equations must be solved numerically.
The density parameters for the components 
of the universe and the effective equations of state for the 
DE and DM for a typical solution are shown in figure \ref{fig:omegas}.
Notice that the transition to the tracker behavior in this example is 
occurring presently.
%
%
%
%

\section{Cosmological constraints and the age of the universe in the exponential 
VAMP model} \label{sec:age}

We have calculated the age of the universe for 
the models that satisfy the Hubble parameter and the 
dark energy density observational constraints
($h=0.72\pm 0.08$, $0.6 \leq \Omega_{\phi 0} \leq 0.8$). 
We have used
stepsizes $\Delta \lambda = \Delta \beta = 0.2$ and $\Delta V_0 = 0.05 
\widetilde{\rho_c}$
for the region $\lambda = [0.01,20]$, $\beta = [0.01,20]$, 
$V_0 = [0.1\widetilde{\rho_c}, 0.8\widetilde{\rho_c}]$, generating $1.4 \times
10^{5}$ models.

Fitting the distribution of models as function of ages, the age of the
universe in this VAMP scenario was found to be  
\begin{equation} \label{eq:vampsage}
t_0 = 15.3 ^{+1.3}_{-0.7} \ \mathrm{Gyr} \ \ \ 68 \% \  \mathrm{C. L.} \ , 
\end{equation}
which is considerably higher than the age of models 
of non-coupled dark energy \cite{wmap,copeland}. 
This seems natural, since in these models
the CDM also has an effective negative equation of state and accelerates
the universe. 
This result is very conservative, since it relies 
only on the well established limits on the Hubble constant and the dark energy
density today. 

\section{Conclusion} \label{sec:conclusion}

VAMP scenario is attractive since it could solve the 
problems of exponential dark energy, giving rise to a solution to the 
cosmic coincidence problem. However, we found that in order to obtain
solutions that can provide realistic cosmological 
parameters, the constant $V_0$ has to be
fine tuned in the range $V_0 = [0.25\widetilde{\rho_c},0.45\widetilde{\rho_c}]$
at 68$\%$ C. L.. This implies that the attractor
is being reached around the present epoch. 
In this sense, the model
is not able to solve the coincidence problem.  

A generic feature of this class of models is that the universe is older than
non-coupled dark energy models. 
Better model independent determination of the age of the universe could help to
distinguish among different contenders to explain the origin of the dark energy. 

\section*{Acknowledgments}
R. R. would like to thank the organizers of the V SILAFAE for their efforts in 
providing a nice atmosphere for the participants and to CNPq for financial
support.

\end{document}